# The Tomaco Hybrid Matching Framework for SAWSDL Semantic Web Services

Thanos G. Stavropoulos, Stelios Andreadis, Nick Bassiliades,
Dimitris Vrakas and Ioannis Vlahavas

**Abstract**— This work aims to resolve issues related to Web Service retrieval, also known as Service Selection, Discovery or essentially Matching, in two directions. Firstly, a novel matching algorithm for SAWSDL is introduced. The algorithm is hybrid in nature, combining novel and known concepts, such as a logic-based strategy and syntactic text-similarity measures on semantic annotations and textual descriptions. A plugin for the S3 contest environment was developed, in order to position Tomaco amongst state-of-the-art in an objective, reproducible manner. Evaluation showed that Tomaco ranks high amongst state of the art, especially for early recall levels. Secondly, this work introduces the Tomaco web application, which aims to accelerate the wide-spread adoption of Semantic Web Service technologies and algorithms while targeting the lack of user-friendly applications in this field. Tomaco integrates a variety of configurable matching algorithms proposed in this paper. It, finally, allows discovery of both existing and user-contributed service collections and ontologies, serving also as a service registry.

——————————— ◆ ———————————

## 1 INTRODUCTION

THE Web currently enables billions of users to look up massive amounts of data and carry out transactions, playing a vibrant role in everyday lives. Since its creation, it has evolved from hosting plain data on static pages to dynamic crowd-sourcing content publication through blogging and social networks. On top of data retrieval, web users can now perform actions and carry out tasks enabled by the Web Service technology. Web Services entail a set of technologies and standards to properly define APIs of remote procedure calls (RPCs). As opposed to Web Applications, Services provide neither a Graphical User Interface (GUI) nor a composite series of actions (but rather atomic actions). On the contrary, Web Services can be invoked by human or software agents to carry out a single atomic operation at a time. Since Web Services need to be discovered before they are invoked, the Semantic Web [1] technologies have been employed to aid that cause. Semantics on service descriptions render them machine interpretable and thus enhance Discovery [2], Selection, Matching and Composition [3].

The Semantic Web Service concepts have emerged from the synergy of Web Service and Semantic Web technologies. Early works in the field followed a top-down approach of describing services using high-level ontological constructs, such as the OWL-S upper ontology for services [4] or the WSMO Web Service Modeling Language [5]. These so-called upper ontologies define hierarchies of concepts for modeling and describing a service, its workflow, grounding and, most importantly, the Input, Output, Preconditions and Effects (IOPEs) of its operations. Ontological descriptions provide much flexibility and expressiveness, but are less strict towards bindings to actually invoke the Services in question, since WSDL groundings are optional. Additionally, their complexity and subjective interpretation has hindered their wide adoption by the industry. This led to the emergence of lightweight, bottom-up approaches, such as SAWSDL [6] and WSMO-lite [7], which provide compact annotations on WSDL groundings themselves. Amongst them, SAWSDL became a W3C recommendation and a leading Semantic Web Service description methodology, which is increasingly adopted in industry and academia. Despite its low complexity and expressiveness, it has already been proven to be suitable for enhancing Web Service Discovery [2], Selection, Matching and Composition [3] [8]. Yet, these notions are just starting to penetrate the real-world with applications, e.g. [9].

This work serves a two-fold contribution towards accessible Semantic Service Matching i.e. the problem of user or software-driven search for suitable services, or in essence service operations, using structured criteria i.e. input and/or output, in accordance to SAWSDL capabilities (which exclude e.g. effects). The first contribution is a proposed Semantic Web Service matching strategy tailored to the SAWSDL lightweight schema. After a thorough review of state-of-the-art algorithms that target SAWSDL and some interesting work on OWL-S/WSMO, most strategies can be classified in categories, such as semantic (logic-based), syntactic (IR-based), structural,

- T. G. Stavropoulos is with the Aristotle University of Thessaloniki, Greece. E-mail: athstavr@csd.auth.gr.
- S. Andreadis is with the Aristotle University of Thessaloniki, Greece. E-mail: sandreai@csd.auth.gr.
- N. Bassiliades is with the Aristotle University of Thessaloniki, Greece. E-mail: nbassili@csd.auth.gr.
- D. Vrakas is with the Aristotle University of Thessaloniki, Greece. E-mail: dvrakas@csd.auth.gr.
- I. Vlahavas is with the Aristotle University of Thessaloniki, Greece. E-mail: vlahavas@csd.auth.gr.

learning and hybrid. The proposed algorithm aims to adopt, adapt and combine some of these elements. The hybrid technique introduced in this work employs a novel logic-based strategy, complemented by text-similarity measures on both semantics and textual descriptions. Evaluation on a large, publicly available dataset has proven that the logic-based technique has the greatest impact on retrieving relevant services. These findings are in-line with works such as [10], where feature vectors extracted from semantic information for retrieval learning, were found more effective than those originating from syntactic information. However, although text-similarity methods give poor results on their own, they do compensate for losses of the logic-based strategy in this work. They also enable semantic matching of non-annotated (plain WSDL) descriptions. Various measures e.g. macro-averaging precision at standard recall levels, F1-score, and average precision (AP) are used to appreciate the effectiveness of each variant. The proposed method has ranked high amongst state-of-the-art in both effectiveness and performance.

The second contribution of this work is the integration of the proposed matching strategies in the user-oriented Tomaco web application (Tool for Matching and Composition). Tomaco is a publicly available web application that aims to render Semantic Web Service exploitation easily accessible to experts and non-experts alike. Additionally to invoking strategies with a variety of parameters to choose from, the users are able to target existing service collections or upload their own. This way, Tomaco also serves as a service registry, allowing discovery using semantic criteria.

This paper is structured as follows: the next section surveys existing state-of-the-art algorithms, datasets and registries, identifying the main principles of matchmaking applications. The third section introduces the Tomaco algorithm, describing its four matching strategies, implementation and extensive evaluation runs. The fourth section presents the Tomaco web application, its architecture and functionality. Future directions and conclusions drawn from this work are presented on the corresponding final section.

## 2 RELATED WORK

This section presents indicative examples of state-of-the-art in matching algorithms, highlighting the underlying principles and techniques. The first subsection focuses on matching systems, which usually do not provide a graphical user interface and are not accessible for use in general. The second subsection presents some examples of web-based registries of services for discovery.

### 2.1 WEB SERVICE MATCHING ALGORITHMS

Existing techniques, as presented on Table 1 and described in the following subsections, can be classified as logic-based, syntactic or text-similarity-based, while structural similarity and learning are less common. Logic-based techniques range from straightforward series of few class-relationship rules to large lists of semantic conditions. Notice that, despite the name, this family of techniques mainly employs reasoning upon class subsumption and equivalence, avoiding more complicated inferences. Text-similarity mostly uses textbook algorithms from the field of Information Retrieval. Each existing technique is ultimately compared to the one proposed in this paper. Measures of effectiveness for most of them are also presented in the evaluation section. When describing logic-based techniques, offered (or provided) and requested services from now on will be denoted as $O$ and $R$ respectively. Input and output are denoted in subscript; e.g. $R_i$ stands for requested input. Superclass and subclass relationships of a left-hand side entity to the right-hand side are denoted using > and < respectively.

### *SAWSDL-MX, SAWSDL-TC, ISEM*

The work in [11] presents two major contributions. First of all, the authors provided the first and largest dataset of more than a thousand SAWSDL files, namely SAWSDL-TC (Test Collection). To do so, they used expert manual labor combined with the OWLS2WSDL tool to map OWL-S descriptions (from OWLS-TC2.2) to lightweight format. The test collection's updated version, SAWSDL-TC3, found online[1], contains 1080 service documents, OWL ontologies, queries and relevant sets. Secondly, the authors present an initial approach to exploit this test collection towards automatic matching. Following the dataset's transformation, SAWSDL-MX is an adaptation of previous works OWLS-MX [12] and WSMO-MX [13].

The algorithm provides all standard matching strategies, namely logic-based, syntactic (text-similarity) and hybrid (logic-based and syntactic similarity). The strategies target service input, output and their underlying components (e.g. ComplexType), trying to find a match between a requested service (i.e. query) and all services offered in a set. Each offered service's operation is matched with every requested operation and rated with the maximum observed match. An offered service's overall rating is the worst (minimum) rating of all requested operations. However, SAWSDL-TC contains single-operation services only. Hence, operation-match rating is equal to the overall service-match rating.

Rating scores for the logic-based strategy are set according to the following order from highest to lowest:
1. *Exact:* perfect matching of inputs and outputs i.e. $R_i = O_i \wedge R_o = O_o$
2. *Plug-in:* offered input is arbitrarily more general than requested input and offered output is a direct child of requested output i.e.
   $O_i > R_i \wedge O_o <_{direct} R_o$
3. *Subsumes*: inputs as in *Plug-in* and offered output is a(ny) child of requested output (relaxes output constraints) i.e. $O_i > R_i \wedge O_o < R_o$
4. *Subsumed-by:* inputs as in *Plug-in* and offered output is a direct parent of requested outputs i.e.
   $O_i > R_i \wedge O_o >_{direct} R_o$

---

[1] SAWSDL-TC online:
http://projects.semwebcentral.org/projects/sawsdl-tc/

5. *Fail:* none of the above applies

When multiple annotations are present, the strategy considers the first one only, while input and output match ratings count equally for the overall service rating. The syntactic strategy applies IR (Information Retrieval) methods i.e. various text-similarity algorithms, provided by the SimPack[2] Java library (e.g. Loss-of-information, extended Jaccard, Cosine and Jensen-Shannon), between requested and offered semantic elements. Finally, the hybrid strategy offers two variants. The *compensative variant* lets syntactic-similarity "compensate" when logic-based returns *Fail*. In the *integrative variant*, *Subsumed-by* matches are further constrained by a text-similarity-above-threshold requirement, and is, thus, more strict than logic-based. The algorithm is integrated along with a Service Registry, Ontology Handlers (which locate ontologies referred in services and host reasoning utilities) and an Ontology Registry. SAWSDL-MX has been evaluated over SAWSDL-TC, indicating the hybrid method (with Cosine text-similarity) as the most effective but also the slowest, followed by the syntactic and logic-based methods.

SAWSDL-MX, being an early approach, offers much room for improvement. First of all, the logic-based approach performed worse than text-similarity. We speculate that this was due to the strict conditions of the former, involving both input and output. Through experiments, we confirmed that independent input/output conditions and looser logic-based conditions are much more effective for the task at hand. Additionally, more effective text-similarity techniques have been found and employed in Tomaco. Finally, SAWSDL-MX picks the lowest of all requested operations ratings per service, which is strict. On the contrary, Tomaco is operation-centric: it rates and returns each individual operation within services, which is less strict.

SAWSDL-MX2 [14], in addition to logic-based and text-similarity, measures structural similarity between WSDL file schema information (e.g. element names, data types and structural properties), using WSDL-Analyzer[3]. It also introduces an adaptive, learning layer where SVM training vectors consist of values for logic-based, textual and structural criteria and binary relevance: {*Exact, Plug-in, Subsumes, Subsumed-by, Fail*, text-similarity, structural-similarity, relevance}. Logic and structural similarity (M0 + WA), adaptive (MX2) and logic + textual hybrid (MX1) show no significant difference on Average Precision (AP), while improving over plain methods. However, M0 + WA and MX2 require double per query response time than MX1. These findings support our decision to employ hybrid logic + textual methods in Tomaco.

iSeM [15] is an evolution of the MX series by the same authors. In principle, it applies SVM learning for the weighted aggregation of underlying algorithm rankings. The learning vectors are an extended version of the ones in –MX2, containing logic, structural and text similarity in similar fashion to –MX algorithms. However, approximate logic matching was added which captures more matches than the existing one, using looser criteria for subsumption. The algorithm is extremely precise in SAWSDL-TC and has ranked first in AP, as measured in S3 (Semantic Service Selection Contest)[4] 2010, 2012. Since the algorithm performs every known technique in state-of-the-art, it comes with an expected trade-off in performance. On the other hand, the proposed Tomaco algorithm targets a fast performance, excluding learning in favor of more confined, straightforward techniques.

Table 1. State-of-the-art comparison

| System | Year | Format | Logic | Syntactic | Hybrid | Other |
|---|---|---|---|---|---|---|
| SAWSDL-MX | 2008 | SAWSDL | yes | yes | yes | - |
| SAWSDL-MX2 | 2009 | SAWSDL | yes | yes | yes | Learn. Struct. |
| iSeM | 2010 | OWL-S, SAWSDL | yes | yes | yes | Learn. Struct. |
| LOG/COM 4SWS | 2010 | SAWSDL | yes | yes | yes | Learn. |
| iMatcher | 2011 | SAWSDL | yes | yes | yes | Learn. |
| Skyline | 2008 | OWL-S | yes | - | - | - |
| HSW | 2009 | SAWSDL | yes | yes | - | - |
| OOM | 2007 | OWL-S | yes | - | - | - |
| OWLS-SLR | 2010 | OWL-S | yes | - | - | - |
| IRS-III | 2008 | WSMO | yes | - | - | - |
| Themis-S | 2010 | WSDL | - | - | - | Lingual |
| WSColab | 2010 | WSDL | - | - | - | Tags |
| URBE | 2009 | SAWSDL | yes | - | yes | Lingual |

As a side-note, we argue that learning in the context of our work could result in loss of generality, due to the lack of additional datasets to test and evaluate the approaches. Generally speaking, learning could be used to select the best strategy for a given instance. As opposed to adaptive works mentioned throughout the paper, Tomaco aims to provide a non-adaptive method for two reasons: there is a lack of training sets to target open-world services and learning would increase the web application's response time, at least without targeted optimizations.

## LOG4SWS, COV4SWS

XAM4SWS is a common framework, which derived two algorithms, LOG4SWS and COV4SWS [16]. Both algorithms perform operation-centric matching, targeting service interfaces, operations and I/O. LOG4SWS performs logic-based matching, in an –MX fashion, mapping ratings to numbers using linear regression. Meanwhile, COV4SWS rating measures are inspired from the field of semantic relatedness. It then performs regression to find weights for the aggregation of ratings (from underlying 4service elements to an overall service rating). Both algorithms fallback to WordNet similarity (inverse distance), if semantics are entirely absent. Both methods are highly effective on TC3, ranking first in nDCG (normalized Discounted Cumulative Gain for graded relevance) and Q-measure [17], while maintaining a fast response time. Tomaco on the other hand, follows a different, non-adaptive approach, as discussed above.

---

[2] SimPack: http://www.research-projects.uzh.ch/p8227.htm
[3] WSDL-Analyzer: http://www.wsdl-analyzer.com/
[4] S3 Contest - http://www-ags.dfki.uni-sb.de/~klusch/s3/

### iMatcher

iMatcher [18] integrates interesting variations of well-known strategies. The first strategy includes three sub-strategies. It performs text-similarity (using Java SimPack) targeting either the WSDL service name field, service description field or semantic annotations. The second strategy selects the maximum rating between two sub-strategies. The first is a hybrid variant where the logic-based part rates inputs and outputs of operations with 1, if the requested concept is a parent of the offered cept ($R > 0$). Hence, iMatcher's logic substantially differs from SAWSDL-MX, which requires $R_o > O_o$ in *Plug-In* and *Subsumes* and the opposite, $R_i < O_i$, in all cases (*Exact*, *Plug-In* etc.). If logic-based matching fails, syntactic matching is performed (in the spirit of the compensative variant). The second sub-strategy examines distance of two concepts originating from different ontologies using ontology alignment similarity as obtained from the Lily tool[5]. On top of that, iMatcher also implements an Adaptive Matching method. The user selects multiple strategies, the results of which form vectors of the training set. Learning is performed by selecting an algorithm from the Weka library. Regarding effectiveness, iMatcher's best strategies are Adaptive Matching with Logistic and ε-SVR learning, followed by hybrid with ontology alignment. However, unlike Tomaco, iMatcher (as SAWSDL-MX2), begins with a less effective, strict logic-based technique that is harder for hybrid methods to improve upon.

### Skyline

The Skyline system [19] performs matching on OWL-S descriptions instead of SAWSDL, but its interesting strategy is worth mentioning. The strategy's target components are IOPEs, grouping Inputs together with Preconditions and Outputs with Effects. First, it performs logic-based classification to *Exact*, *Direct_Subclass*, *Subclass*, *Direct_Superclass*, *Superclass*, *Sibling* and *Fail* (selecting the best match). The homonymous Skyline algorithm is used to find the optimal trade-off of input versus output significance. E.g. a service of *Exact* input and *Fail* output and a second of *Subclass* input and *Direct_Superclass* output will prevail over a third service of *Direct_Subclass* input and *Fail* output. Ratings for multiple $R_i, R_o$ are also supported through the Skyline algorithm. Users are able to request the next skylayer from the algorithm to get more services. Skyline (0.83 AP) has ranked well above OWLS-MX (0.71 AP) on OWLS-TC2 dataset.

---
[5] Lily linguistic and structural ontology alignment tool : http://ontomappinglab.googlepages.com/oaei2007

The Skyline technique has actually proven to be effective in the document retrieval domain. However, to adapt it in service matching, the authors had to consider semantic relationships as ordinal values, e.g. a superclass is worse than a subclass. Our approach does consider that some relationships are superior to others but this case differs for input and output concepts. Overall, the Skyline technique does seem interesting, if properly adapted, and may be investigated as future work.

### Hybrid Strategy with WordNet

The SAWSDL matching system in [20], denoted by HSW, proposes a complex Logic-based algorithm to classify Input and Output as *Precise*, *Over*, *Partial*, *Mismatch* according to different ratios of provided, matched and requested I/O. The algorithm entails a long series of rules for the classification (which arguably makes its practical meaning hard to grasp). E.g. when semantics are missing from the examined node but the parent node's semantics match, text-similarity and WordNet [21] distance on the examined node are used as a measure. No evaluation was performed to assess the system's effectiveness.

While WordNet seems to be a popular choice regarding similarity measures, we feel that the presented algorithm's criteria are too strict i.e. the logic-based strategy entails a complex series of conditions to be met for matching. The text-based and WordNet-based approaches also entail hard-to-meet criteria (i.e. missing semantics on the current node but similar semantics on the parent node) which does not allow them to compensate for the logic-based strategy. For that purpose, we propose a compensative hybrid technique to handle mismatches that indeed improves over pure strategies. Furthermore, WordNet is a lexicon itself, while service requests always come with their own lexicons, i.e. ontologies to serve as heuristics. However, it can be used as a semantic and syntactic measure all-in-one in a future endeavor.

### Object-Oriented Measures and OWLS-SLR

The work in [22], denoted as OOM, proposes a novel method for measuring similarity in Semantic Web Service Matching. The algorithm maps OWL-S to COOL (CLIPS Object Oriented Language) and considers Simple (i.e. data type) and Relational (i.e. logic-based) Property similarity for OWL-S signatures (I/O). Three categories are considered for Simple Property: *exact match*, *numerical-type match* (e.g. xsd:int and xsd:float) and *mismatch*. The logic-based variant in OOM is called Relational Property similarity and is equal to the distance between two classes in a hierarchy or through a common ancestor. If no common ancestor exists or classes are disjoint, their distance is infinite and, thus, similarity is zero. The overall Relational rating is the average of ratings per multiple $R_i, R_o$ also using weights. Finally, total rating is equal to the product of Simple and Relational scores.

The Simple Property similarity idea in this work seems intuitive, but essentially turns out to serve numeric type targets only, which are rare. Likewise, the common ances-tor distance in Relational Property similarity is an appealing idea to relax search criteria, but at the expense of lower precision and longer reasoning time. Finally, this system does not handle the absence of semantics.

An evolution of OOM, is the OWLS-SLR system [23], which considers semantic relationships of both subsumption and siblings. It targets OWL-S profile I/O signatures, looking into ontology properties (roles) in addition to classes. The strength of OWLS-SLR is its low response time, as it focuses on finding fast an initial set of candidate descriptions. On the contrary, our work does not consider such advanced reasoning capabilities, as precision is improved by a straightforward logic-based and syntactic strategies combined.

### IRS-III

IRS-III [24] is an integral system for WSMO creation, execution and selection, but also works with OWL-S through an import mechanism. A custom ontology representation, OCML, is used to encode service descriptions. The selection/matching subsystem, namely Goal Mediator, selects Web Services that match the requested capabilities, which range from input types, preconditions and assumptions to non-functional properties. Mismatches fall under two cases: either requested (Goal) inputs are different or fewer in number than the ones offered i.e. $R_i \neq O_i \vee |R_i| < |O_i|$. A usage scenario demonstrates an eGovernment application about agencies and emergency planning. A more recent version of the system took part in the S3 contest in 2009, where the underlying logic-based variant used OCML rules to match I/O concepts. It ranked low during JGD (Jena Geography Dataset) experiments, on both effectiveness with an average precision of 0.41 and performance of 2.826s per query. The system's logic-based strategy is unclear and thus cannot be evaluated. Additionally, using ad hoc fields and criteria in service descriptions, such as assumptions, and non-functional properties is considered a non-universal practice.

### THEMIS-S

The system presented in [25] focuses on syntactic matching, applying no logic-based method. The so-called enhanced Topic-Based Vector Space Model (eTVSM), a variant of classic TVSM, is extracted from WSDL descriptions. This variant uses WordNet to appreciate linguistic relations between natural language terms and classify matches as *synonymy*, *homonymy*, *hyponymy* or *hypernymy*. For evaluation purposes, the authors constructed their own dataset consisting of a hundred service descriptions about geo Web Services, written in English natural-language text form (74 to 1271 words each) and derived from programmableweb.com[6] and seekda.com[7]. Domain experts constructed suitable queries and relevant sets for two different scenarios of thirty queries each. The proposed algorithm, eTVSM, outperformed state-of-the-art in this context but performed moderately during the S3 contest. Although the use of WordNet for text-similarity is a promising method, our work mainly demonstrates the superiority and effectiveness of logic-based methods.

### WSC<sub>OLAB</sub>

The authors of [26] have come up with the novel idea of utilizing collaboration and social web principles e.g. tags on service annotation and retrieval. They have developed a portal that enables users to individually annotate services by manually providing tags. These tags define either service behavior (i.e. categorization of service functionality), input and output service interface or identification of additional characteristics. Queries are manually formulated by experts as well, using a web portal. The portal also provides auto-complete suggestions from the already known service tags during that process. Finally, matching is performed by constructing a Vector Space Model (VSM) of behavior, input and output tags. The system participated in the JGD track of S3 2009, outperforming all algorithms on both average precision (0.54) and performance (~0s). While the presented collaborative filtering approach seems very promising for offline tagging of services, it presents certain issues for online matching. Query transformation, from natural language to the tag terms, is human-driven, introducing subjectivity and hindering online matching. In other words, while users are able to quickly search the repository by using tags, services cannot be dynamically tagged and indexed.

### URBE

The work in [27] proposes the URBE/URBE-S system, which incorporates a hybrid SAWSDL matching algorithm. The logic-based strategy, named URBE-S, calculates *annSim* (annotation similarity) as the distance of two concepts in the same ontology (as in [22]). If semantics are non-existent (pure URBE case), *nameSim* finds linguistic similarity, targeting service name, operation name and I/O, using a domain-specific or general purpose ontology, such as WordNet in this implementation. In both cases, *DataTypeSim* calculates data type similarity between xsd:types in WSDL simpleTypes (only) according to a predefined table. Overall rating is set to the average of $R_i, R_o$ ratings. Macro-averaging precision-recall diagram ranked URBE-S above plain URBE and various state-of-the-art algorithms in [27]. All in all, the URBE/URBE-S system, despite its long response time, justifies the effectiveness of semantics in matching algorithms.

## 2.2 WEB SERVICE REGISTRIES

Some past works have been more focused on providing service registries rather than effective matching algorithms. OPOSSum [28] is such a web-based registry of services, that also hosts large datasets e.g. JGD and OWLS-TC. However, the underlying retrieval technique is a keyword-based mapping to SQL queries and not based on semantics. A similar, but much more extensive effort can be seen in BioCatalogue [9], a large registry of services related to life science. BioCatalogue similarly does not support semantic queries, but extends keyword search with much more metadata search fields and a tagcloud. A similar example is WESS [29], a keyword-search service registry that discovers WSDL/SAWSDL and OWL-S files after targeted crawling over the Web. Finally, iServe [30] follows a different service description approach, using RDF/Linked Data to publish, analyze and discover services. Compared to such works, we do provide a much more concise service registry, but focus on semantic search and matching. In other words, Tomaco provides an algorithm for individuals to experiment with semantic matching algorithms and parameters on existing or user-provided service collections.

---

[6] Repository of Web APIs: http://www.programmableweb.com/
[7] Repository of bookings: https://www.seekda.com/

## 3 TOMACO MATCHING ALGORITHM

This section presents the matching algorithm introduced in this work. Its four underlying strategies are explained in detail, the latter of which is a hybrid combination of the former three. The next subsections present the description of the strategies, algorithm implementation details and an extensive effectiveness and performance evaluation.

### 3.1 TOMACO MATCHING STRATEGIES

*LOGIC-BASED STRATEGY*

The proposed logic-based strategy collects semantic annotations from target components of the XML-based tree structure of SAWSDL/WSDL. Algorithm 1 shows the outline of the complete rating strategy. Instead of providing pseudocode for the extraction procedure (invoked at *op.getInput*, line 4 and *op.getOutput*, line 12), Fig 1 shows the standard WSDL structure to be explored in a recursive DFS manner, while collecting *sawsdl:modelReferences* when present. The algorithm begins from *interface* and goes down to tree leaves (usually *simpleTypes*) exhaustively (choosing any branch first, e.g. between input or output, type or element). Specifically, the unique service *interface* may have one or more *operations*, which in turn, have *input* and *output*. *Input* and *output*, through *message*, have *parts* or *elements*. Each *part* can contain a *type* or, again an *element*. *Types* can be either *simpleTypes* or *complexTypes*. The latter can contain a sequence *(xsd:sequence)* of *elements* and so it can infinitely continue to decompose to simpler types. In this proposed strategy, the depth of annotation is considered irrelevant. In other words, semantics on e.g. a *part*, direct child of *input*, or a *simpleType*, on a long nesting of *types* and *elements*, are practically of the same significance.

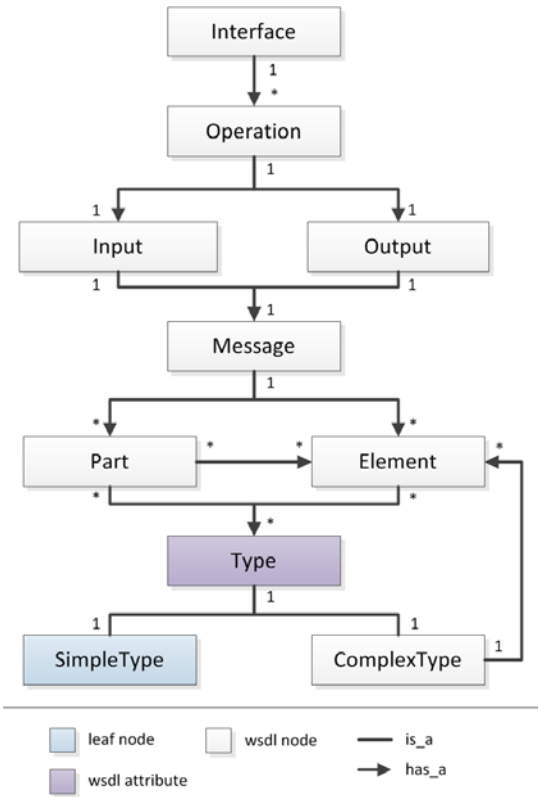

**Fig 1. WSDL tree definition that guides DFS-traversal for the extraction of semantics**

The logic-based strategy of rating semantics themselves, handles characteristics of inputs and outputs differently, as presented in Algorithm 2. More specifically, it is based on a practical principle that, in the context of I/O-driven search, users may possess certain input information and/or desire certain output information. Hence, they can settle for more abstract input data, i.e. input offered is a superclass of required ($O_i > R_i$), but more specific input data ($O_i < R_i$) is less desired. The opposite of this principle applies for service outputs. The user desires to obtain certain output information from a service. If this information is more generic ($O_o > R_o$), it is of lesser use to the user, while more specific information ($O_o < R_o$) is of greater use. Some examples within the SAWSDL-TC dataset that conform to these principles are Genre > Science_Fiction (books.owl), MedicalOrganization > Hospital (HealthInsuranceOntology.owl) for input and City < UrbanArea (travel.owl), OpticalZoom < Zoom (extendedCamera.owl) in desired outputs.

Overall, the algorithm is driven by intuitively identifying a user's *desires*. We consider four matching cases: *Exact*, *Desired*, *LessDesired* and *Fail*. The *Exact* match is the most desired one, in both cases of input and output, and should be rated with maximum similarity. To rate the rest of the cases, two parameters are defined, an *UpperRate* and a *LowerRate*. Hence, when a *Desired* concept is offered, i.e. a superclass of input or a subclass of output, logic-based similarity is set to *UpperRate*. Likewise, if a *LessDesired* concept is found, i.e. a subclass of input or a superclass of output, rating is set to *LowerRate*. If no semantics are present or the concepts share no hierarchical relationship, the match is classified as *Fail*. The proposed method always handles similarity as a numerical value in [0, 1] in order to allow continuous values of ratings and facilitate combinations with other methods (e.g. hybrid). Through internal experiments, we have found the optimal values to be 0.75 for *UpperRate* (close but less than *Exact*) and 0.25 for *LowerRate* (close but higher than *Fail*).

- *Exact* : 1

**Algorithm 1** Tomaco Matching

**Input:**
$R_i$: Array of requested input;
$R_o$: Array of requested output;
$O_s$: Set of offered services;
$w$: weight of input vs output significance, $\in (0,1)$
**Output:** Ranking of offered operations

```
1: Array OperationRatings ← ∅;
2: for each service s ∈ O_s
3:   for each operation op ∈ s.getOperations()
4:     Array O_i ← op.getInput().getSemantics();
5:     Array ratingO_iR_i = ∅      //max per r_i ratings
6:     for each requested_input r_i ∈ R_i
7:       Array ratingO_ir_i ← LogicBased(r_i, O_i,'input');
8:       ratingO_iR_i.add(ratingO_ir_i.getMax());
9:     end for
10:    double ratingO_i ← ratingO_iR_i.getAverage();
11:
12:    Array O_o ← op.getOutput().getSemantics();
13:    Array ratingO_oR_o = ∅      //max per r_o ratings
14:    …        //as in ratingO_oR_o
15:    double ratingO_o ← ratingO_oR_o.getAverage();
16:
17:    //apply weights for overal operation rating
18:    if R_i ≠ ∅ ∧ R_o ≠ ∅
19:      ratingOp ← w * ratingO_i + (1 − w) * ratingO_o;
20:    else if R_i = ∅
21:      ratingOp ← ratingO_o;
22:    else
23:      ratingOp ← ratingO_i;
24:    end if
25:    OperationRatings.add(ratingOp);
26:  end for
27: end for
28: return OperationRatings.sort(desc);
```

**Algorithm 1. The Tomaco matching function performing DFS extraction of semantics, logic-based rating and applying weights**

- *Desired* : UpperRate = 0.75
- *LessDesired* : LowerRate = 0.25
- *Fail* : 0

The strategy entails a series of getting max values, average of vectors and applying weights between input and output, as shown on Algorithm 1. In detail, matching is performed by rating each offered operation. Two vectors are formed, one for requested inputs (line 5) and one for requested outputs (line 13). Each vector value is the maximum rating score between the requested concept and all underlying offered concepts (lines 6-9). Consequently, the overall input and output ratings per offered operation are the average values of the corresponding vectors (lines 10, 15), to provide balance. Actually, other algorithms get the minimum instead of the average of vector values, but this is intuitively stricter and as seen during internal experiments with TC3, provides worse results. The final offered operation rating is the weighted sum of input and output ratings (lines 17-23).

Note here, that this different handling of input and output does not mean that their importance is different. Input versus output significance is a separate parameter given in the form of weight (line 18). Usually, this is set to

**Algorithm 2** Logic Based

**Input:**
$r_x$: requested input;
$O_x$: Array of requested output;
$x$: input or output identifier;
**Output:** Array of $O_x$ ratings according to $r_x$
**Constants:**
$UpperRate$: high rating for desired matches e.g. 0.75;
$LowRate$: low rating for less desired matches e.g. 0.25;

```
1: Array ratingO_xr_x ← ∅;
2: for each offered o_x ∈ O_x
3:   if o_x ≡ r_x
4:     rating ← 1;     //Exact
5:   else if o_x > r_x    //superclass of r_x
6:     if x = input
7:       rating ← UpperRate;   //Desired case
8:     else
9:       rating ← LowerRate;   //Less desired case
10:    end if
11:  else if o_x < r_x    //subclass of r_x
12:    if x = output
13:      rating ← UpperRate;   //Desired case
14:    else
15:      rating ← LowerRate;   //Less desired case
16:    end if
17:  else
18:    rating ← 0;    //Fail
19:  end if
20:  ratingO_xr_x.add(rating);
21: end for
22: return ratingO_xr_x;
```

**Algorithm 2. The Tomaco logic-based rating strategy**

0.5 unless a user desires otherwise. Additionally, the proposed algorithm does not apply weights in case one of the two, input or output, is entirely not requested (lines 20-23). On the contrary, when input or output is requested but does not match (*Fail*), overall rating is reduced accordingly.

Finally, the algorithm does not further normalize per operation rating for service ratings. Instead, each operation is returned as a whole. This approach is based on the principle that service operations are self-contained methods of certain input and output, whereas services are containers of such methods. In other words, the algorithm essentially performs operation matching and is able to handle operations independently. A rating threshold can also be applied as filtering means to improve quality versus quantity of retrieved services. This threshold was not applied during evaluations, for completeness, but is rather suitable for the web application system.

### SYNTACTIC-ON-SEMANTICS STRATEGY

The Syntactic-on-Semantics (denoted as Syn-On-Sem) method's purpose is to compensate for mismatches of the logic-based method i.e. when classes are not related but their names are similar. Typically, the Syn-On-Sem strategy handles semantic annotations as plain textual expressions, applying text-similarity metrics. In other words, the strategy measures syntactic text-similarity between re-quested semantics and offered semantic annotations.

First of all, the algorithm itself obtains semantic annotations using the same DFS-traversing strategy as the previous method. However, this time, the so-called unfolding of semantics occurs, meaning the annotation string URI is trimmed before the '#' character to just get a class name. Consequently, $R_i, R_o$ and $O_i, O_o$ are matched to provide an operation rating for the requested input and output. The core text-similarity matching is provided by a library of textbook methods. Note that pseudocode is not provided for compactness, but rather explained here. The aforementioned changes occur just by substituting the logic-based method calls in Algorithm 1 by a suitable method for the unfolding and measuring text similarity.

After experimenting and analyzing different text-similarity methods, we selected the two methods most suitable in this context. Semantic names, like variable names in programming, follow different conventions to comply with the disallowance of spaces. For this reason, standards such as CamelCase, i.e. capitalization of the first letter of each word in a phrase, and snake_case, i.e. underscore between words, were long ago established in programming. Incidentally, CamelCase is the dominant naming convention in SAWSDL-TC as well. As a result, target words to be found in text are in fact substrings.

For such string instances, the algorithms Monge-Elkan [31] and Jaro [32] were found to be more suitable in respect to others e.g. Cosine, Dice, Euclidean Distance, Jaccard, and Levenshtein. Their high suitability is justified, as Monge-Elkan was especially designed to match atomic strings, where words are delimited by special characters, while Jaro targets short strings such as names. In internal experiments, Monge-Elkan was found to rate desired, similar instances of CamelCase strings high (not lower than 1), as did Jaro (above 0.7). Hence, we have set the corresponding thresholds for a text-similarity match at 1 and 0.7 respectively.

### SYNTACTIC-ON-SYNTACTICS STRATEGY

The Syntactic-on-Syntactics (denoted as Syn-On-Syn) strategy completely disregards semantics on offered input and output, performing text-similarity on the names of target WSDL elements. Hence, it is the only fruitful strategy to perform when semantic annotations are few or absent (plain WSDL files). The algorithm is rather similar with the previous strategies except minor changes in the pseudocode of Algorithm 1. Instead of collecting semantics, the tree is again traversed collecting each element's name string. Those names are then compared to each $r_i$ and $r_o$ forming an overall rating and ranking of operations. The text-similarity algorithms employed here are again Monge-Elkan and Jaro, since syntactic description element names are again expressed in CamelCase, snake_case or similar conventions.

### HYBRID STRATEGY

The final proposed strategy is hybrid in nature and employs the other three strategies roughly in an order of effectiveness. As all the aforementioned strategies examine different targets and even use different heuristics, they can compensate for one another. The effectiveness of each method is highly dependent on the subjective definition of relevance per query. Experiments in SAWSDL-TC (Section 3.3) have shown that the logic-based method is the most effective amongst pure strategies, and is, thus, used in highest priority.

The proposed hybrid method intermediates syntactic matching between the *Exact* and non-*Exact* logic-based matches. On the contrary, hybrid methods in literature mainly exhaust logic-based matching up to *Fail*, before exploring textual-similarity (e.g. the compensative method in [11]). Furthermore, in Tomaco, Syn-On-Sem and in turn Syn-On-Syn are performed after an *Exact* mismatch. The rating of syntactic measures (Monge-Elkan or Jaro) in case of a match is 1, just as in the *Exact* case. The principle behind this choice is that, instinctively, if there is a syntactic similarity match (i.e. rating above per algorithm threshold), the offered concept is exactly the one sought; not its parent, child or sibling. Hence, the rating should reflect the one of *Equal* match (1). If even text-similarity fails, the *Desired* and *LessDesired* logic checks occur resulting in lower ratings. All in all, the method reinforces the chances of true matches to get a high, *Exact*, rating.

We believe that our approach gives a higher chance for syntactic matching to compensate for *Exact* false negatives (high rating and checking before fail). This claim is supported in the evaluation section where Tomaco performs higher than other hybrid methods. Notably, we chose for the syntactic method to compensate for logic-based, since the latter performs much better than the former out-of-the-box. The opposite would be counter-intuitive in this context, where logic matches are higher. From another perspective, one could choose or learn which method to use in each case (see adaptive algorithms). However, in the scope of this work, we are investigating for a straightforward non-adaptive method to target service matching.

## 3.2 ALGORITHM IMPLEMENTATION

The proposed algorithm has been implemented entirely in Java, as were all underlying libraries and tools used. To explore the XML-like WSDL structure, instead of using a pure XML general-purpose parser, we used easyWSDL [33], an SAWSDL-specific library. The library provides modelReferences for each WSDL element while traversing the structure in a DFS manner. However, minor issues do exist. Attributes (e.g. modelReference in this case) could not be extracted from *wsdl:operation*, while wsdl:type attribute could not be retrieved from wsdl:part, hindering DFS path. In both cases, we had to manually retrieve, parse and identify attributes to resolve these issues.

For the logic-based method particularly, an ontology parsing tool and a reasoner had to be employed. After retrieving ontology references from the offered descriptions, the OWL-API Java tool [34] and its underlying reasoner, Hermit [35] are employed to evaluate their rela-tionships with requested concepts. Reasoning upon the ontology allows the system to answer queries such as if the requested class is equivalent, more general or more specific than the offered class. OWL-API and Hermit are invoked for the same tasks during both Logic-based and Hybrid strategy and remain idle during the rest. For strategies that entail text-similarity, we used the Simmetrics Java library[8] that provides a wide selection of suitable algorithms. As mentioned before, numerous internal evaluation runs were performed to obtain the optimal text-similarity algorithms and thresholds for the context of semantic service descriptions.

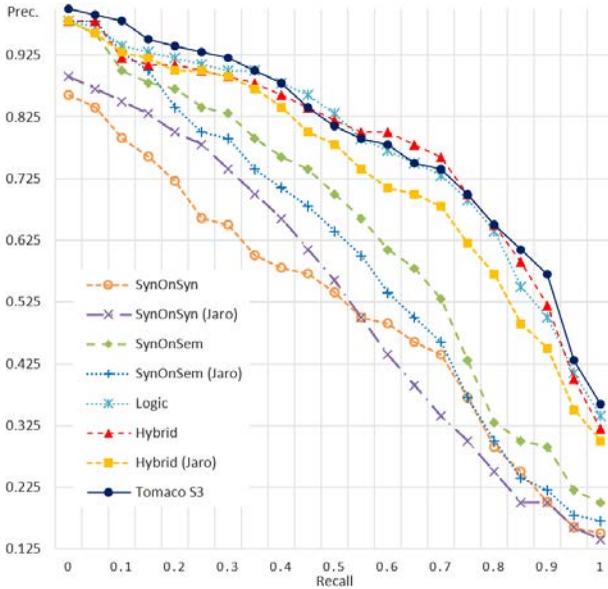

**Fig 2. Macro-averaging precision of Tomaco variants**

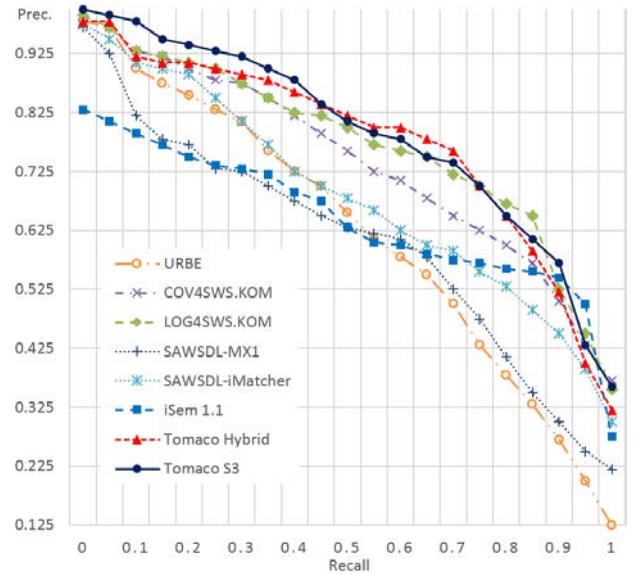

**Fig 4. Macro-averaging precision of Tomaco Hybrid and Tomaco-S3 amongst S3 2012 participants**

Performance-wise, some technical developments were oriented towards optimizing the algorithm's response time. First of all, reasoners are known to introduce long delays due to the complexity involved in such tasks. To resolve this matter, an initialization is performed loading all reasoners at start-up, in order to answer a batch of queries on-the-fly (e.g. all queries in the evaluation run). Secondly, traversing the service descriptions introduced the longest delay. For this reason, syntactic and semantic extraction and indexing of all terms in an internal structure (hash table) is also performed during initialization, to allow on-the-fly comparisons. Code optimizations resulted in one of the fastest response times amongst state-of-the-art, presented in the next section. Note that this code optimization was tailored to the experiments while a more suitable one is incorporated in the integrated Tomaco web application, as presented in the following section.

### 3.3 ALGORITHM EVALUATION

SAWSDL-TC3 [11] is one of the most eminent, uniform and large SAWSDL datasets publicly available for evaluation purposes. On top of that, its authors conduct the yearly S3 contest which provides an integrated evaluation platform both for precision and performance of all participants. Although, the Tomaco algorithm was not designed with the particular dataset or contest in mind, the S3 environment appears as a suitable opportunity to evaluate Tomaco against state-of-the-art and guarantees objectivity when measuring various metrics. Therefore, we developed the Tomaco plugin[9] for the SME2 contest environment[10], which readers can use to reproduce the results. The plugin contains all proposed Tomaco variants. We also added the Tomaco-S3 variant, which specifically targets the contest. Tomaco-S3 syntactically compares query names with operation service names before performing the Tomaco hybrid method (rating matches higher than hybrid matches). Tomaco-S3 improves rank-ings but it is naturally excluded from the open-ended Tomaco web application, as query names are not provided in user-defined queries as in S3. The plugin was also submitted to the S3 2014 call as a byproduct of this work. An additional byproduct was the release of a modified TC3 with syntactic corrections in some ontologies, which caused problems, at least with the libraries used in Tomaco. The set is available online[9] and under examination by the TC3 authors.

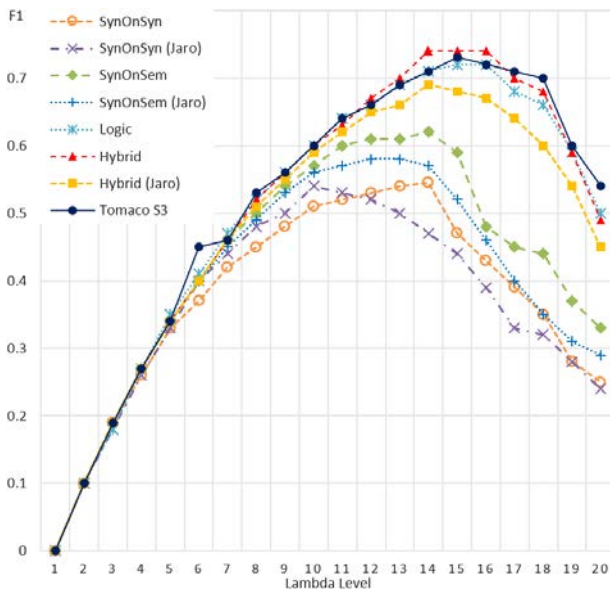

**Fig 3. F-measure of Tomaco variants**

[8] SimMetrics: http://sourceforge.net/projects/simmetrics/

[9] Tomaco homepage, offering the SME2 plugin and TC3 modification: http://lpis.csd.auth.gr/people/thanosgstavr/applications/tomaco.html

[10] The SME2 tool: http://projects.semwebcentral.org/projects/sme2/

Table 2. Measures and performance of proposed Tomaco variants in SME2

|  | Logic-based | Syn-On-Syn | Syn-On-Syn (Jaro) | Syn-On-Sem | Syn-On-Sem (Jaro) | Hybrid | Hybrid (Jaro) | Tomaco-S3 |
|---|---|---|---|---|---|---|---|---|
| AP | 0.767 | 0.507 | 0.522 | 0.633 | 0.586 | 0.771 | 0.725 | **0.785** |
| nDCG | 0.851 | 0.594 | 0.67 | 0.749 | 0.742 | 0.838 | 0.826 | **0.868** |
| Q | 0.772 | 0.506 | 0.54 | 0.633 | 0.605 | 0.761 | 0.741 | **0.795** |
| AQRT (s) | 0.131 | 0.315 | 0.041 | 0.07 | **0.029** | 0.245 | 0.137 | 0.376 |
| Total (m) | 0.414 | 0.549 | **0.344** | 0.386 | 0.346 | 0.541 | 0.422 | 0.593 |

Table 3. Measures and performance of Tomaco and S3 2012 participants in SME2

|  | Tomaco-S3 | Tomaco Hybrid | iSeM 1.1 | LOG4SWS | COV4SWS | Nuwa | iMatcher | URBE | SAWSDL-MX1 |
|---|---|---|---|---|---|---|---|---|---|
| AP | 0.785 | 0.771 | **0.842** | 0.837 | 0.823 | 0.819 | 0.764 | 0.749 | 0.747 |
| nDCG | 0.868 | 0.838 | 0.803 | **0.896** | 0.884 | 0.884 | 0.855 | 0.85 | 0.839 |
| Q | 0.795 | 0.761 | 0.762 | **0.851** | 0.825 | 0.817 | 0.784 | 0.777 | 0.767 |
| AQRT (s) | 0.376 | 0.245 | 10.662 | **0.241** | 0.301 | 9.009 | 1.787 | 40.01 | 3.859 |

To adapt Tomaco variants to the contest we set a rating threshold of zero, in order to return a ranking for all offered services. The input-vs-output weight was set to 0.5, since the organizers state that I/O carries the same significance. Textual similarity employs Monge-Elkan (threshold = 1) by default and Jaro (threshold = 0.7) when mentioned so. Overall, TC3 is compatible with the Tomaco process. It contains 1080 SAWSDL documents from nine domains: education, medical care, food, travel, communication, economy, weapons, geography and simulation. Each service has a single interface of a single operation, which suits the operation-centric ranking of Tomaco. *ModelReferences* to OWL2-DL ontologies, are placed in *wsdl:parts* and underlying elements. The set includes 42 predefined queries for the contest in SAWSDL form and respective relevant sets in XML form. 38 OWL ontologies are used within services and queries alike.

Initially, we used the SME2 tool to measure metrics and performance of the proposed Tomaco variants. Macro-averaging precision at twenty recall levels (Fig 2) and F1 score on twenty lambda levels (Fig 3) both show the superiority of the S3 and Hybrid variants. The former exceeds overall, especially in early levels, while the latter exceeds in mid-levels. Table 2 shows that the same ranking holds for AP, while for nDCG and Q the logic-based variant ranks second, after S3 and before Hybrid. All measures dictate that, although pure textual methods perform much lower than logic-based, they do improve performance of the latter when combined in the Hybrid method. Additionally, the S3 variant improvement over Hybrid, shows that query names indeed play a significant role in TC3. Internally, we also experimented with the impact of reasoning for semantically equivalent classes as *Exact* matches, which indeed improved logic and Hybrid methods (by 2% and 1% AP respectively). The experiments also show the significance of textual-similarity algorithm selection, as Jaro is found less accurate for the most part. We also internally experimented with other algorithms which significantly lowered AP (e.g. Levenshtein by 53% on Syn-On-Syn).

Table 4. Time decomposition for Tomaco variants in (s)

|  | Logic-based | Syn-On-Syn | Syn-On-Sem | Hybrid |
|---|---|---|---|---|
| Total time | 60.887s | **37.861s** | 44.320s | 71.370s |
| Init. Reasoners | 11.638s | **0.000s** | 0.000s | 11.466s |
| DFS extraction | 40.809s | **35.209s** | 35.318s | 39.499s |
| All queries | 8.440s | **2.652s** | 9.002s | 20.405s |
| Per query avg. | 0.201s | **0.063s** | 0.214s | 0.486s |

Performance-wise, the Jaro variants are faster than Monge-Elkan, while compromising precision. Meanwhile, the most accurate variants S3 and Hybrid manage to sustain a reasonably fast per query average response time (AQRT) and a total running time of around half a minute. Additional experiments were targeted to break total time down, outside the SME2 tool (which slightly alters running time). Table 4 shows total time decomposition in reasoner initialization (if applicable), DFS-extraction, rating for all 42 queries and per query response time. Total query time for Syn-On-Sem is notably higher than for Syn-On-Syn, surprisingly enough, due to unfolding. All experiments were performed on an Intel i5 @3.20GHz, 8.00GB RAM.

Table 3 presents the two most effective variants, S3 and Hybrid, amongst state-of-the-art algorithms as presented on the S3 contest of 2012[4] [17], using the same tool and dataset. Tomaco variants rank after iSeM, LOG4SWS, COV4SWS and Nuwa [17] in AP, but above iMatcher, URBE and SAWSDL-MX1. In nDCG and Q-measure, only LOG4SWS, COV4SWS and Nuwa surpass Tomaco-S3. Meanwhile, Tomaco variants perform significantly better in macro-averaging precision at recall levels, shown on Fig 4. Tomaco-S3, especially, ranks above other algorithms for the most part, except the final couple of levels, where other algorithms prevail. Hereby we conclude that the proposed algorithms are optimal when a portion, e.g. top-k, relevant services are required and not all of them (as a side-note, top-k AP is currently not available for comparison neither in SME2 nor in literature).

Additionally, the published per query response times rank Tomaco just after LOG4SWS, COV4SWS, hence, showing a fair trade-off of precision versus computational time. Please note that we did not re-run other algorithms (since some of them are not available online) neither do we know the specs of the S3 2012 environment, so performance is not directly comparable but it is reported here only for the sake of completeness. Apparently, learning algorithms take significantly more time, with the exception of iMatcher and the top ranking ones. We also speculate that learning algorithms also require significant time for initialization i.e. training (please note that total times and technical specs of the S3 contest are not available at the time).

As a general remark, Tomaco does not rank first in most metrics, but shows an optimistic performance in the macro-averaging precision graph. It is also apparent that it performs exceptionally well for a large percentage of recall levels. Hence, it can be useful for use cases where the user demands most, but not all, relevant documents while maintaining a satisfactory response time.

## 4   TOMACO WEB APPLICATION

The Tomaco web application is an integrative Tool for Matching and Composition of Web Services available on the Web[11]. While matching is the main issue discussed in this work, composition constitutes an additional issue in the Web Service lifecycle, entailing different motivation, problems and solutions. Hence, only the matching counterpart is presented here. This section states the motivation behind designing and developing the Tomaco web application, its functionality, software architecture and technical implementation.

The motivation behind this attempt is twofold. First of all, to provide Web Service developers, researchers and consumers with a ready to use algorithm for matching their own, as well as existing Semantic Web Services, in a user-friendly graphical manner. Secondly, through ease-of-use and online availability, Tomaco aims to advertise and accelerate the uses of Semantic Web Services, SAWSDL/ WSDL and the Semantic Web altogether. This point is especially important since the lack of user-friendly and functional tools for Semantic Web technologies (as acknowledged in [36]) is an eminent area for improvement.

### 4.1 TOMACO APPLICATION ARCHITECTURE

Fig 5 presents an abstract layout of Tomaco's application architecture. Users are able to access the application's Web Graphical User Interface through any web browser. Through it, they can add files to the Ontology and Service repositories stored at the server. Secondly, they can form queries using ontology terms on the GUI. Consequently, the matching subsystem provides strategies to match the query across Service storage and return a filtered ranking of results to the GUI. The embedded reasoner is used during logic-based and hybrid strategies within the matching engine. Likewise, a text-similarity algorithm library is used during Syntactic and Hybrid matching strategies.

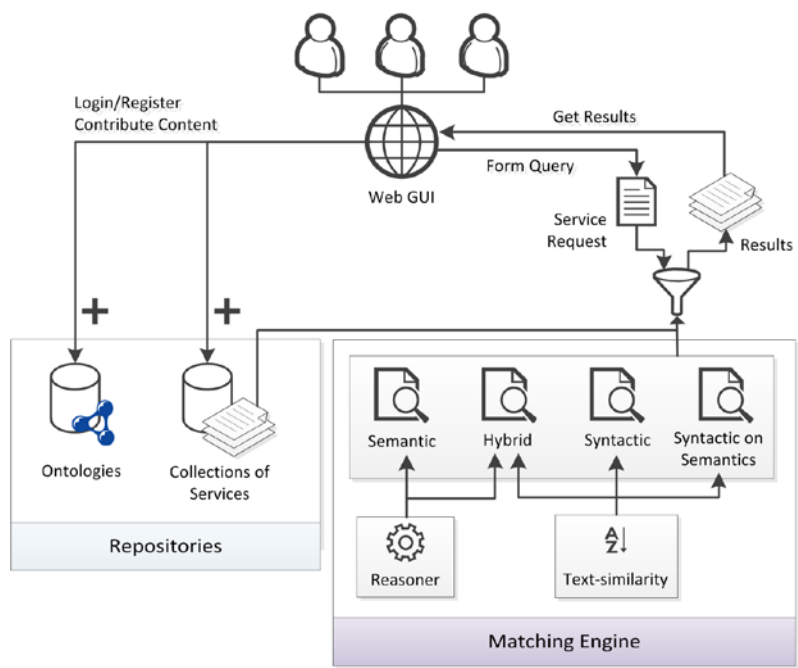

**Fig 5. The Tomaco Web Application infrastructure, underlying repositories, matching engines and user interaction**

[11] Tomaco web application: http://tomaco.csd.auth.gr

## 4.2 FUNCTIONALITY

The first major purpose of Tomaco is to provide a comprehensive registry of Web Services, organized in collections. Collections mostly reflect the origin of Web Services, e.g. SAWSDL-TC 3 constitutes a collection, although services within it target different application domains. Users are able to browse existing service repositories of large or small collections or register to upload their own. New services can be uploaded from local user stores or linked from online sources. Certain metadata such as the uploader, date and text description can be entered and viewed for each collection. Service names can be searched using a list with Autocomplete. Upon selecting a service, its description is displayed in a tree hierarchy. The purpose of having collections is for them to serve as different search frontiers for the application's algorithms. Similarly to Services, Tomaco stores a collection of ontologies, necessary for forming queries. The Ontologies section allows users to browse or add ontologies from local files or link them from online sources. The collection can be browsed using an Autocomplete list. When selecting an ontology, its metadata and its contents, classes and their hierarchy are displayed.

The second section of the application provides algorithms for automatic service matching and composition. The matching process integrates the algorithm proposed in this work (Section 3). Users firstly have to select one of the service collections as the algorithm's search frontier i.e. offered services. In order to construct a query, i.e. provide $R_i$ and/or $R_o$, users have to select these concepts from the available ontologies by browsing them in a graphical manner. Numerical parameters to provide include the input-vs-output weight and rating threshold both in the range of 0.1 to 0.9. A rating threshold is introduced here to retrieve more refined results.

Users can select a matching method, choosing one of the four strategies proposed in this work: Logic-based, Syn-On-Sem, Syn-On-Syn or Hybrid. For methods that entail text-similarity measures, both Monge-Elkan and Jaro methods are provided (although Monge-Elkan has been proved to be the best choice so far, this may not be the case in other datasets). Matching results are presented on a table ordered in descending rating order. The table displays each operation's name, rating, service and interface that it belongs to. The user can finally view a justification for the rating of each result, namely, the name, type and semantics of the underlying matched offered element with the highest rating. A comprehensive example of query formation, configuration and retrieval is depicted in Fig 6.

## 4.3 SYSTEM IMPLEMENTATION

The Tomaco application is currently hosted on a single Apache Tomcat server. The server hosts service and ontology file copies locally for constant availability; even when users link online services or ontologies, those files are retrieved and stored on disk. However, file upload metadata, such as descriptions, are stored in a MySQL database along with user profiles. WSDL files are parsed and indexed at upload time to boost retrieval times. The matching subsystem employs all technologies and libraries already used for the algorithm implementation detailed before. These include easyWSDL for parsing service descriptions, OWL-API for traversing and Hermit for reasoning upon ontologies and Simmetrics for invoking various text-similarity algorithms. The GUI was implemented using HTML, CSS, JavaScript and jQuery. Different functions are performed using JSP, e.g. retrieving local files, or JavaScript/jQuery to invoke servlets for func-tions. Java servlets are used to retrieve service/ontology metadata from the database, to retrieve and copy uploaded files and to invoke the matching subsystem.

**Fig 6. Web Service matching query formation and retrieval in Tomaco's graphical user interface**

In order to optimize real-time performance, Tomaco invokes DFS-extraction and indexing of service elements in the database, on service-upload time (instead of a hash table in memory when running experiments in SME2). Although it is not practical to construct reasoners for an arbitrary number of ontologies in working memory, as in the experiments, per query response time of the web application remains low (in the order of a few seconds).

## 5 CONCLUSIONS AND FUTURE WORK

This paper introduces a Semantic Web Service matching algorithm for SAWSDL, entailing three pure strategies, namely logic-based, Syntactic-On-Syntactics and Syntactic-On-Semantics, and a hybrid composite strategy. It also presents the integration of the algorithm in a web application named Tomaco, along with a service registry and an architecture to provide on-demand matching, performed on both existing and user-contributed content. The underlying algorithm's evaluation is carried out using the S3 contest environment, which allows reproducing results and positions Tomaco amongst state-of-the-art algorithms. The proposed logic-based method proves more effective than pure text-similarity strategies. However, text-similarity, with appropriate algorithms, does significantly improve and compensate for logic mismatches in the proposed hybrid variant. Meanwhile, Tomaco ranks high amongst state-of-the-art algorithms, especially for the initial recall levels i.e. when a portion of relevant services is required. Optimizations, such as indexing and preprocessing, resulted in a satisfactory low response time for the Tomaco web application.

Future work is mainly focused towards two directions: enriching the algorithm itself and extending the Tomaco system. The strategies proposed in this work can possibly benefit from future developments in information retrieval, such as the Google or Flickr distance metrics, which have yet to be explored in the context of service retrieval. Future variants could also examine more fine-grained ontological similarity e.g. concept distance and additional text-similarity targets e.g. rdfs:label, rdfs:comment (which are not rich enough in current datasets). From another perspective, since queries and relevancy are subjective, voting techniques can be explored to ensure the system selects the optimum strategy in various cases. On enriching the Tomaco web application, we plan to focus on defining motivation and innovative methods towards user-accessible service composition. Technically, its overall usability can be improved by providing a REST API for invoking the system's functions and allow software agents to discover the services, while exploring further integration with existing service registries and providers. Finally, collecting feedback and community interaction are critical towards improving of the platform.

## 6 ACKNOWLEDGMENTS

The authors wish to thank Thodoris Tsompanidis for his earlier work, Dr. Efstratios Kontopoulos, Dr. George Meditskos and the reviewers for their fruitful comments and the S3 organizers for their work in the field.

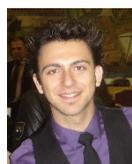

**Thanos G. Stavropoulos** is a PhD Student at the Department of Informatics at the Aristotle University of Thessaloniki (AUTH), after receiving his BSc (2009) and MSc (2011) degrees from the same department. Meanwhile, he has worked in projects such as Dem@Care FP7 (at the Centre for Research and Technology Hellas) and Smart IHU (at the International Hellenic University). His research interests include intelligent autonomous systems, semantic web services, ambient intelligence and sensor networks, while he is Chair of the ACM Student Chapter of AUTH. [http://users.auth.gr/athstavr]

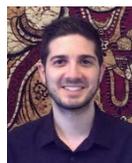

**Stelios Andreadis** has received his BSc (2011) and MSc (2014) degrees in Information Systems, both from the Department of Informatics of the Aristotle University of Thessaloniki (AUTH), receiving scholarships for excellence. His research interests include Semantic Web Services, Information Retrieval and Machine Learning, while he is also Treasurer of the ACM Student Chapter of AUTH.

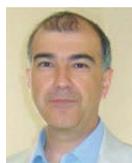

**Nick Bassiliades** received his MSc degree from the Computing Science Department of Aberdeen University (1992) and his PhD degree in parallel knowledge base systems from the Department of Informatics of the Aristotle University of Thessaloniki (1998), where he currently serves as an Associate Professor. His research interests include knowledge-based and rule systems, multi-agent systems, ontologies, linked open data and the Semantic Web. He has published more than 150 papers in journals, conferences, and books, and coauthored two books, receiving over 1000 citations (h-index 17). He has been involved in 30 R & D projects and has been co-Chair e.g. in RuleML and WIMS, and in the Program Committee for more than 70 conferences/workshops. He is director of RuleML, Inc. and member of the Greek Computer Society, IEEE, and ACM. [http://tinyurl.com/nbassili]

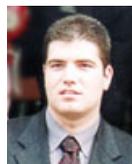

**Dimitris Vrakas** is a Lecturer at the Department of Informatics at the Aristotle University of Thessaloniki, Greece, after receiving his PhD degree in Intelligent Planning Systems from the same department in 2004. His research interests include artificial intelligence, intelligent tutoring systems, planning, heuristic search and problem solving. He has published more than 50 papers in journals and conferences, and co-edited 2 books. He has been involved in projects concerning intelligent agents, e-learning and web services [http://lpis.csd.auth.gr/vrakas].

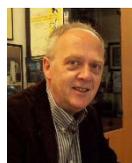

**Ioannis Vlahavas** is a Professor at the Department of Informatics of the Aristotle University of Thessaloniki, after receiving his Ph.D. degree in Logic Programming Systems from the same department (1988). He specializes in machine learning, knowledge-based and AI systems and has published over 250 papers and 9 books, receiving 3700 citations (h-index 30). He has been involved in more than 30 projects and served as chair in many conferences [http://www.csd.auth.gr/~vlahavas].